\documentclass[10pt,journal,compsoc]{IEEEtran}

\usepackage{amsmath,amsfonts}
\usepackage{algorithmic}
\usepackage{array}
\usepackage{textcomp}
\usepackage{url}
\usepackage{verbatim}
\usepackage{graphicx}

\usepackage[dvipsnames]{xcolor}
\usepackage[most]{tcolorbox}
\usepackage{multirow}
\usepackage{graphicx}
\newcommand{\ie}{\emph{i.e., }}
\newcommand{\eg}{\emph{e.g., }}

\newcommand{\zjz}[1]{{#1}}
\newcommand{\bkq}[1]{{#1}}
\newcommand{\zy}[1]{{#1}}
\newcommand{\revision}[1]{{#1}}

\newcommand{\aka}

\usepackage{enumitem}
\usepackage{verbatim}
\usepackage[ruled]{algorithm2e}
\usepackage{multirow}
\usepackage{subfigure} 
\usepackage{ragged2e}
\usepackage{caption}
\usepackage{graphicx}
\usepackage[normalem]{ulem}
\useunder{\uline}{\ul}{}
\usepackage{amsmath}
\usepackage{setspace}
\usepackage{bm}
\usepackage{longtable}
\usepackage[numbers,sort&compress]{natbib}

\def\BibTeX{{\rm B\kern-.05em{\sc i\kern-.025em b}\kern-.08em
    T\kern-.1667em\lower.7ex\hbox{E}\kern-.125emX}}
\usepackage{balance}
\begin{document}
\title{ CoLLM: Integrating Collaborative Embeddings into Large Language Models for Recommendation}

\author{Yang Zhang, Fuli Feng, Jizhi Zhang, Keqin Bao, Qifan Wang and Xiangnan He
\thanks{This work is supported by the National Key Research and Development Program of China (2022YFB3104701) and the National Natural Science Foundation of China (U24B20180, 62121002). (\textit{Corresponding authors: Fuli Feng; Xiangnan He}.) }
\thanks{Yang Zhang, Fuli Feng, Jizhi Zhang, and Keqin Bao are with the University of Science and Technology of China, China.  Email: zyang1580@gmail.com, fulifeng93@gmail.com, cdzhangjizhi@mail.ustc.edu.cn, baokq@mail.ustc.edu.cn.}
\thanks{Qifan Wang is with Meta AI, USA. Email: wqfcr618@gmail.com.}
\thanks{Xiangnan He is with the MoE Key Lab of Brain-inspired Intelligent Perception and Cognition, University of Science and Technology of China, China. Email: xiangnanhe@gmail.com.}
}


\markboth{Journal of \LaTeX\ Class Files,~Vol.~18, No.~9, September~2020}%
{How to Use the IEEEtran \LaTeX \ Templates}

\maketitle

\begin{abstract}
Leveraging Large Language Models as recommenders, referred to as LLMRec, is gaining traction and brings novel dynamics for modeling user preferences, particularly for cold-start users. However, existing LLMRec approaches primarily focus on text semantics and overlook the crucial aspect of incorporating collaborative information from user-item interactions, leading to potentially sub-optimal performance in warm-start scenarios. To ensure superior recommendations across both warm and cold scenarios, we introduce \textit{CoLLM}, an innovative LLMRec approach that explicitly integrates collaborative information for recommendations. CoLLM treats collaborative information as a distinct modality, directly encoding it from well-established traditional collaborative models, and then tunes a mapping module to align this collaborative information with the LLM's input text token space for recommendations. By externally integrating traditional models, CoLLM ensures effective collaborative information modeling without modifying the LLM itself, providing the flexibility to adopt diverse collaborative information modeling mechanisms. Extensive experimentation validates that CoLLM adeptly integrates collaborative information into LLMs, resulting in enhanced recommendation performance. Our implementations are available in Github: 
{\url{https://github.com/zyang1580/CoLLM}}.
\end{abstract}


\section{Introduction}
\label{sec:into}
\IEEEPARstart{L}{arge} Language Models (LLMs) like GPT3~\cite{instructGPT,GPT3} and LLaMA~\cite{llama} have made rapid advancements, showcasing remarkable capabilities in context comprehension, reasoning, generalization, and modeling world knowledge, among others~\cite{zhao-survey}. These exceptional proficiencies have sparked intense interest and enthusiasm for exploring and utilizing LLMs across diverse fields and disciplines~\cite{blip2, singhal2023large, singh2023progprompt,10417790}. 
Recommender systems, as a core engine for personalized information filtering on the web,
are also anticipated to reap significant benefits from the development of LLMs. For instance, the world knowledge and context comprehension abilities of LLMs could enhance item understanding and user modeling, particularly for cold items/users~\cite{LLMrec-cold}. This anticipation opens up an exciting new direction: leveraging LLMs as recommenders (LLMRec)~\cite{ehcheng-survey, qing-llmrec}, which exhibits the potential to become a transformative paradigm for recommendation~\cite{tallrec,ehcheng-survey, zhang2023chatgpt}.

\begin{figure}[t]
	\centering
	\includegraphics[width=0.49\textwidth]{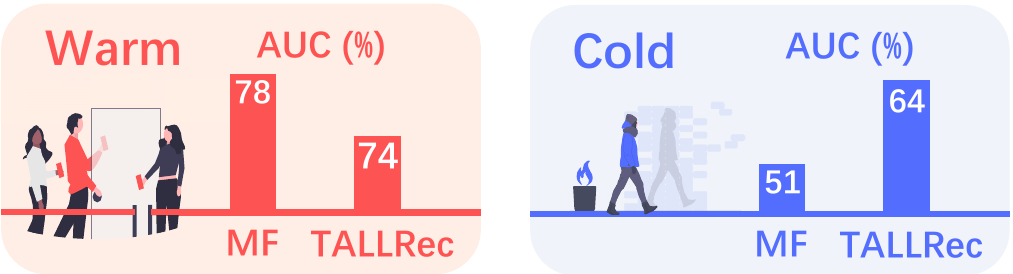}
    \caption{A demonstration of LLMRec method (TALLRec~\cite{tallrec}) performance compared to conventional methods (MF~\cite{mf}) in warm and cold scenarios on Amazon-Book~\cite{amazonbook} data.}
	\label{fig:warm_and_cold_intro}
\end{figure}

To leverage LLMs as recommenders, pioneering studies have relied on In-Context Learning~\cite{GPT3}, which involves directly asking LLMs to make recommendations by using natural language-based prompts~\cite{LMRecSys, uncoveringGPTRec, chat-rec, llmsKnowPref}. However, most empirical findings indicate that the original LLMs themselves struggle to provide accurate recommendations, often due to a lack of specific recommendation task training~\cite{bigrec, instructfollows,tallrec}. To address this challenge, increasing efforts have been devoted to further fine-tuning LLMs using relevant recommendation data~\cite{bigrec, instructfollows, tallrec, llmsKnowPref}. Nevertheless, despite incorporating tuning to learn the recommendation task, these methods could still fall short of surpassing well-trained conventional recommender models, particularly for warm users/items, as demonstrated in recent works~\cite{rella} and Figure~\ref{fig:warm_and_cold_intro}.

We argue that the primary limitation of existing LLMRec methods is their inadequate modeling of local collaborative information implied within the co-occurrence patterns in user-item interactions.
These methods represent users and items using text tokens, relying predominantly on text semantics for recommendations, which inherently fall short of capturing collaborative information. 
For example, two items with similar text descriptions might possess distinct collaborative information if consumed by different users, yet this difference often goes unaccounted for due to the textual similarity. 
Nevertheless, collaborative information between users and items often proves beneficial for recommendations, especially for ones with rich interactions (\textit{i.e.,} warm ones)~\cite{ctrl}. 
\revision{Ignoring such information can lead to suboptimal performance.} 
Hence, we introduce a novel research problem: 
how can we efficiently integrate collaborative information into LLMs to optimize their performance for both warm and cold users/items?

To solve the issue, we propose explicitly modeling collaborative information in LLMs. 
Drawing from the experience of classic collaborative filtering with latent factor models (\eg Matrix Factorization)~\cite{mf,lewu-survey, 9601264}, a straightforward solution is introducing additional tokens and corresponding embeddings in LLMs to represent users/items, akin to the roles played by user/item embeddings in latent factor models. 
\bkq{This enables the possibility of encoding collaborative information when using these embeddings to fit interaction data, similar to MF. 
However, directly adding token embeddings would decrease scalability for large-scale recommendations and increase LLMs' tokenization redundancy, resulting in a lower overall information compression rate~\cite{LLMisCompression}. }
This reduced compression rate is particularly significant considering that collaborative information is typically low-rank, and it could ultimately make prediction tasks (including recommendation) more challenging for LLMs~\cite{LLMisCompression,acl21best}. Moreover, this method lacks the flexibility to incorporate more advanced modeling mechanisms, such as explicitly capturing high-order collaborative relationships like LightGCN~\cite{lightgcn}.

In this light, to effectively enhance LLM with collaborative information in a lightweight and flexible manner, we propose \textit{CoLLM}, a new method that treats collaborative information as a separate modality and introduces it into LLM by directly mapping it from a (well-trained) conventional collaborative model using a Multilayer Perceptron (MLP). CoLLM employs a two-step tuning procedure: first, fine-tuning LLM in the LoRA manner using language information exclusively to learn the recommendation task, and then specially tuning the mapping module to make the mapped collaborative information understandable and usable for LLM's recommendation via considering the information when fitting recommendation data. 
By aligning knowledge from the conventional model with LLMs, CoLLM effectively incorporates collaborative information into LLMs. 
This approach maintains scalability comparable to the original LLMs while also providing flexibility in implementing various collaborative information modeling mechanisms by adapting the choice of conventional models.

The main contributions are summarized as follows:
\begin{itemize}[leftmargin=*]
    \item We highlight the significance of incorporating collaborative information modeling into LLMs for recommendation, so as to make LLMRec perform well in both warm and cold users/items.
    \item 
    We introduce CoLLM, a novel method that effectively integrates collaborative information into LLMs by harnessing the capability of external traditional models to capture the information.
    \item We conduct extensive experiments on two real-world datasets. Extensive results demonstrate the effectiveness of our proposal.
\end{itemize}

\section{Related Work}
In this section, we first discuss the related work on LLMRec. Subsequently, given our focus on integrating collaborative information into LLMs as an additional modality, we would briefly discuss the related work on multimodal LLM.


\subsection{LLMRec}
Recently, with the remarkable emergence of LLMs, there has been a gradual exploration of integrating these sophisticated models with recommender systems~\cite{ llmrecsurvey2, qing-llmrec, ai2023information, ehcheng-survey}.
Some researchers employ the methodology of In-context Learning, relying on the natural language comprehension and generation capabilities of LLMs for recommendation purposes~\cite{chat-rec}.
Among them, Chat-Rec~\cite{chat-rec} facilitates the recommendation process by harnessing the conversational capabilities of ChatGPT. 
Besides, researchers are also endeavoring to facilitate the acquisition of recommendation capabilities in LLMs through in-context learning approaches~\cite{uncoveringGPTRec, ChatgptGoodRec}.
However, because the objective of LLM pre-training is not geared towards recommendation, these methods often exhibit suboptimal performance.
To alleviate this issue, some researchers have employed instruction tuning using empirical recommendation data to enhance the recommendation capabilities of the LLM~\cite{tallrec, instructfollows}, and achieved commendable performance.
Additionally, ~\cite{llmconversationrec, rectool} either employed fine-tuning techniques or utilized prompting on the LLM to enable it to acquire proficiency in using diverse tools for facilitating conversational recommendations.

Although the above works demonstrate the feasibility of tuning LLMs using recommendation data, they still fall short in certain settings compared to traditional models~\cite{rella}.
This can be attributed to the fact that LLMs often heavily rely on semantic priors and tend to overlook collaborative information~\cite{icldiff}.
\revision{To our knowledge, BIGRec~\cite{bigrec} is the only work that addresses this issue beyond ours. However, BIGRec tackles the problem by ensembling LLMs with collaborative models, rather than integrating collaborative information into the LLM generation process, which limits the full potential of LLMs.
Besides, two concurrent works~\cite{jundong,lc-rec} explore the use of collaborative embeddings; however, they focus on directly learning ID embeddings in the LLM space. In contrast, we focus on mapping collaborative embeddings into the LLM space.}
When extending to the field of language models (LMs) for recommendation, some works have concentrated on combining LMs with collaborative models~\cite{ctr_bert, ctrl}. 
Typically, they use the LM's semantic information as a feature for the collaborative model. However, these approaches may face issues related to forgetting semantic information. In contrast, our approach centers around LLM and still relies on the LLM itself to seamlessly integrate semantic and collaborative information, rather than the reverse.


\subsection{Multimodal LLM}
\revision{Among the progress in the field of LLM}, the endeavors that resonate most closely with our work involve the exploration of multimodal LLMs~\cite{vision1,minigpt4,speech1,palme}.
For instance, MiniGPT4~\cite{minigpt4} combines a frozen visual encoder with a frozen advanced language model, revealing that aligning visual features with large language models enables advanced multi-modal capabilities like generating detailed image descriptions and creating websites from hand-drawn drafts.
Palm-E~\cite{palme} aims to integrate real-world continuous sensor inputs into language models, creating a connection between words and sensory information. This integration enables the development of embodied language models capable of addressing robotic challenges.
These works aim to leverage the robust generation and comprehension capabilities of LLM to process textual data and map information from other modalities such as vision and audio to the textual modality, thereby achieving a large multimodal model with language as its primary carrier, which is similar to our motivations.

\section{PRELIMINARIES}
\label{sec:pre}
To begin with, we briefly present the problem definition, the basic concepts of LLMs, and the collaborative models used in our framework.

\par \textbf{Problem Definition}.
Let $\mathcal{D}$ denote the historical interaction dataset. Each data point within $\mathcal{D}$ is represented as $(u, i, y)$, where $u$ and $i$ correspond to a user and an item,  respectively, with $y\in\{1,0\}$ indicating the interaction label. 
Furthermore, there is additional textual information available for items, primarily in the form of item titles. In this study, we explore the utilization of both the interaction data and textual information to fine-tune an LLM for recommendation purposes. Our goal is to enable the LLM to effectively leverage collaborative information beyond text information, achieving superior performance in both warm and cold recommendation scenarios.

\par \textbf{Large Language Model}.
LLMs refer to a class of language models equipped with at least several billion parameters and trained on massive text datasets, showcasing remarkable emergent capabilities~\cite{zhao-survey}. LLMs demonstrate strong proficiency in general natural language understanding and generation, as well as various other aspects such as world knowledge modeling, enabling them to excel at handling a wide range of complex tasks as long as they can be described in language. Typically, LLMs process input text through the following two key steps: 1) tokenization and embedding lookup: in this step, the input text is transformed into meaningful lexical tokens, which are then embedded into a vector space; 2) contextual modeling and output generation (LLM Prediction): LLMs utilize their neural networks, primarily based on a decoder-only transformer architecture, to process the token embeddings obtained in the previous step, generating coherent and contextually relevant output. In this work, we use Vicuna-7B~\cite{vicuna2023} 
for recommendation.

\par \textbf{Conventional Collaborative Recommender}. We mainly consider the latent factor models, such as MF and LightGCN, for encoding collaborative information. 
These approaches typically represent users/items using latent factors, also known as embeddings. Subsequently, they form latent user/item representations through various operations, \eg neighborhood aggregation in LightGCN, to better model collaborative information. Formally, for each sample $(u,i,y)\in \mathcal{D}$, 
\begin{equation}\label{CF-encode}
    \bm{u} = f_{\psi}(u;\mathcal{D}); \quad \bm{i} = f_{\psi}(i;\mathcal{D}), 
\end{equation}
where $\bm{u}\in \mathcal{R}^{1\times d_{1}}$ denotes the user's representation with dimension $d_{1}$, $f_{\psi}(u;\mathcal{D})$ denotes the process used to obtain this representation, similarly for $i$, and $\psi$ denotes model parameters.
The user and item representations are then fed into an interaction module to generate predictions.  By minimizing the prediction errors against the actual interaction labels, the latent representations would learn to encode collaborative information within the interaction data.
\section{METHODOLOGY}
\begin{figure*}[h]
  \centering
  \includegraphics[width=0.85\linewidth]{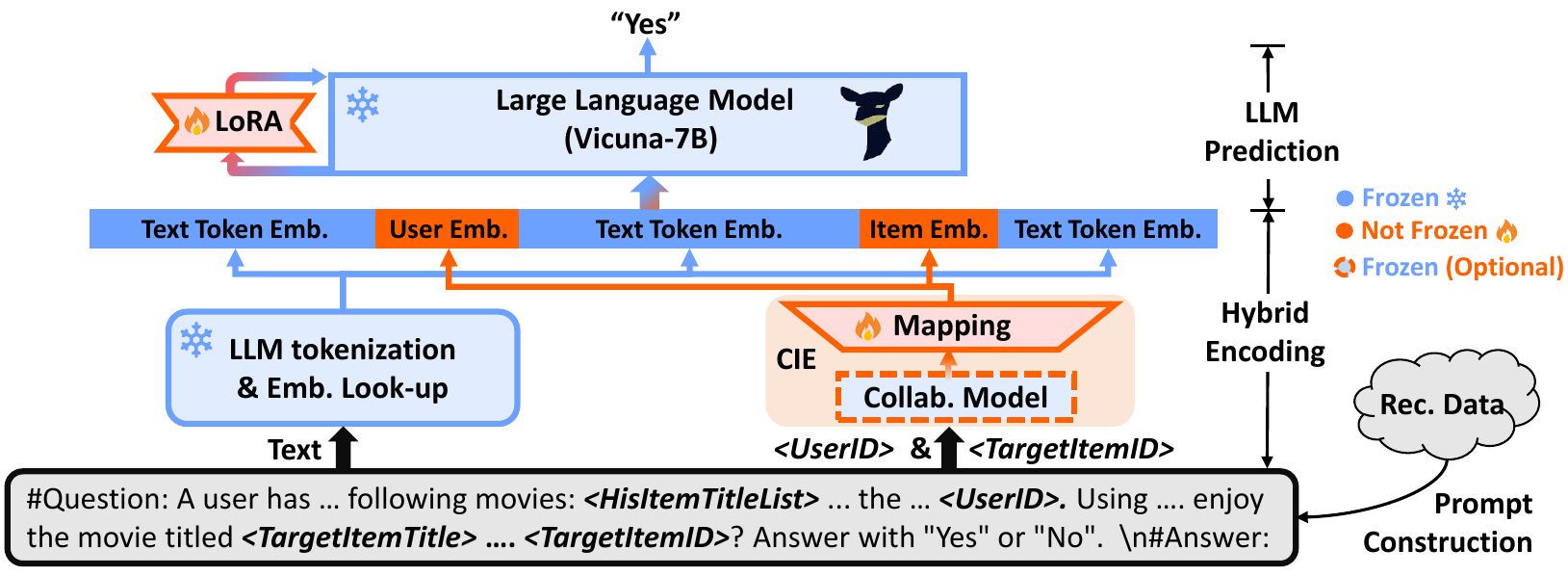} %
\vspace{-5pt}  
  \caption{
  Model architecture overview of CoLLM, comprising three key components: prompt construction, hybrid encoding, and LLM prediction.
  ``Collab.'' is short for ``collaborative'', ``Rec.'' for ``recommendation'', and ``Emb.'' for ``embedding''.
  }
  \label{fig:collabrec}
\end{figure*}

Collaborative information can be viewed as a distinct information modality, capturing user and item co-occurrence relationships in the interaction data. The LLM itself lacks a dedicated mechanism to extract this modality beyond text modality. 
To overcome the limitations, rather than modifying the LLMs directly, we continue to extract collaborative information using conventional models and then transform the extracted results into a format that the LLM can comprehend and utilize for recommendations,
inspired by recent advancements in multimodal LLMs~\cite{blip2,minigpt4}.
This concept forms the foundation of our CoLLM methods. In the following, we provide a detailed overview of CoLLM, beginning with a description of the model architecture designed to connect the conventional models and LLM. We then outline the training strategy that enables the effective integration of collaborative information into LLMs. 







\subsection{Model Architecture}
Figure~\ref{fig:collabrec} illustrates the model architecture of CoLLM, which consists of three components: prompt construction, hybrid encoding, and LLM prediction. Similar to existing approaches, CoLLM starts by converting recommendation data into language prompts (prompt construction), which are then encoded and inputted into an LLM to generate recommendations (hybrid encoding and LLM prediction). Differently, our approach introduces an innovative aspect by incorporating collaborative information to enhance the LLMs' recommendations. This is achieved through the specific designs in the two former components:
\begin{itemize}[leftmargin=*]
    \item  When constructing prompts, we add user/item ID fields in addition to text descriptions to represent collaborative information.
    \item When encoding prompts, alongside the LLMs' tokenization and embedding look-up for encoding textual information, we employ a conventional collaborative model to generate user/item representations that capture collaborative information, and map them into the token embedding space of the LLM.
\end{itemize}

After representing textual and collaborative information within the token embedding space, the LLM could leverage both types of information to perform recommendations. Next, we delve into the specific details of each component.

\subsubsection{Prompt Construction}\label{sec:prompt-template} 
We utilize fixed prompt templates for prompt generation. 
Similar to TALLRec~\cite{tallrec}, we describe items using their titles and describe users by the item titles from their historical interactions. Uniquely, in order to incorporate collaborative information, we introduce additional user and item ID-related fields that do not carry meaningful semantics but serve as placeholders for the collaborative information within the prompt. Ultimately, our fixed prompt template is structured as follows:



\begin{center}
\fcolorbox{black}{gray!6}{\parbox{0.45\textwidth}{\noindent$\bullet$ \textbf{Prompt template.} \#Question: A user has given high ratings to the following items: $\langle$\textit{HisItemTitleList}$\rangle$. \uline{Additionally, we have information about the user's preferences encoded in the feature $\langle$\textit{UserID}$\rangle$}. Using all available information, make a prediction about whether the user would enjoy the item titled $\langle$\textit{TargetItemTitle}$\rangle$ \uline{with the feature $\langle$\textit{TargetItemID}}$\rangle$? Answer with ``Yes'' or ``No''. \#Answer:}}
\end{center}

In this template, ``$\langle$\textit{HisItemTitleList}$\rangle$'' represents a list of item titles that a user has interacted with, ordered by their interaction timestamps, serving as textual descriptions of the user's preferences. ``$\langle$\textit{TargetItemTitle}$\rangle$'' refers to the title of the target item to be predicted.
The ``$\langle$\textit{UserID}$\rangle$'' and ``$\langle$\textit{TargetItemID}$\rangle$'' fields are utilized to incorporate user and item IDs, respectively, for injecting collaborative information. To maintain semantic coherence while integrating user/item IDs, we treat them as a type of feature for users/items within the prompt, as indicated by the underlined content in the template. 
For each recommendation sample, we populate the four fields with the corresponding values of the sample to construct the sample-specific prompt.

\subsubsection{Hybrid Encoding} 

The hybrid encoding component is utilized to convert the input prompt into latent vectors, 
\ie embeddings
suitable for LLM processing. In general, we employ a hybrid encoding approach. As shown in Figure~\ref{fig:collabrec}, for all textual content, we make use of the LLM's built-in tokenization and embedding mechanism to convert it into tokens and subsequent token embeddings. In contrast, when dealing with the ``$\langle$UserID$\rangle$'' and ``$\langle$TargetItemID$\rangle$'' fields, we leverage a Collaborative Information Encoding (CIE) module built with a conventional collaborative recommender, aiming at extracting collaborative information for the LLM to utilize.

Formally, for a prompt corresponding to the sample $(u,i,y) \in \mathcal{D}$, we initiate the process by using the LLM Tokenizer to tokenize its textual content. The tokenization result is denoted as $P = [t_1, t_2, \dots, t_k, u, t_{t+1}, \dots, i, \dots, t_K]$, where $t_k$ represents a text token, and $u$/$i$ signifies the user/item (ID) placed within the ``$\langle$UserID$\rangle$''/ ``$\langle$TargetItemID$\rangle$'' field. We then further encode the prompt into a sequence of embeddings $E$:
\begin{equation}\label{eq:hybird}
    \begin{split}
        E=[\bm{e}_{t_1},\dots,\bm{e}_{t_k},\bm{e}_{u}, \bm{e}_{t_{k+1}}, \dots,\bm{e}_{i}, \dots, \bm{e}_{t_K}],
    \end{split}
\end{equation}
where $\bm{e}_{t_{k}} \in \mathcal{R}^{1 \times d_{2}}$ denotes the token embedding for $t_{k}$ in the LLM with dimension $d_{2}$, obtained via embedding lookup, \ie $\bm{e}_{t_{k}} = Embedding_{LLM}(t_k) $; while $\bm{e}_{u} /\bm{e}_{i} \in \mathcal{R}^{1\times d_{2}}$ denotes the collaborative information embeddings (\ie collaborative embeddings) for the user $u$/item $i$, obtained via the following CIE module.

\vspace{+3pt}
\par \textbf{CIE module}: 
The CIE module consists of a conventional collaborative model ($f_{\psi}(\cdot)$ in Equation~\eqref{CF-encode}) and a mapping layer $g_{\phi}(\cdot)$ parameterized by $\phi$, to extract collaborative information for LLM usage. When provided with the user $u$ and item $i$, the conventional collaborative model generates user and item ID representations ($\bm{u}$ and $\bm{i}$) encoding collaborative information. Subsequently, the mapping layer maps these representations to the LLM token embedding space, creating the final latent collaborative embeddings ($\bm{e}_{u}$ and $\bm{e}_{i}$) used by the LLM.
Formally, we have:
\begin{equation}\label{eq:CIE}
    \begin{split}
       & \bm{e}_{u} = g_{\phi}(\bm{u}), \quad \bm{u} = f_{\psi}(u;\mathcal{D}),  \\
       & \bm{e}_{i} = g_{\phi}(\bm{i}), \quad \bm{i} = f_{\psi}(i;\mathcal{D}),
    \end{split}
\end{equation}
where $\bm{u}=f_{\psi}(u;\mathcal{D}) \in \mathcal{R}^{1\times d_{1}}$ denotes the user representation  obtained by $f_{\psi}$ following Equation~\eqref{CF-encode} for $\mathcal{D}$, similarly for $\bm{i}$. The collaborative model can be implemented as any conventional collaborative recommender as described in Section~\ref{sec:pre}. As for the mapping layer, we implement it as a Multilayer Perceptron (MLP), maintaining an input size equal to the dimension $d_1$ of $\bm{u}/\bm{i}$ and an output size equal to the LLM embedding size $d_2$ (usually $d_{1}<d_{2}$).

\subsubsection{LLM Prediction} 
Once the inputted prompt has been converted into a sequence of embeddings $E$ (in Equation~\eqref{eq:hybird}), the LLM can utilize it to generate predictions. However, due to the absence of specific recommendation training in LLM, instead of relying solely on the LLM, we introduce an additional LoRA module~\cite{lora} to perform recommendation predictions, as depicted in Figure~\ref{fig:collabrec}. The LoRA module entails adding pairs of rank-decomposition weight matrices to the original weights of the LLM in a plug-in manner for specifically learning new tasks (recommendation) while just introducing a few parameters. Then,
the prediction can be formulated as follows:
$$
\hat{y} = h_{\hat{\Theta}+\Theta^{'}}(E), 
$$
where $\hat{\Theta}$ denotes the fixed model parameters of the pre-trained LLM $h(\cdot)$, and $\Theta^{'}$ denotes the learnable LoRA parameters for the recommendation task. 
$\hat{y}$ represents the prediction probability for the label being $1$, \ie the likelihood of answering ``Yes'' for LLM.
The consideration of using LoRA here is that
with the plug-in approach, we only need to update the LoRA weights to learn the recommendation task, enabling parameter-efficient learning.

\subsection{Tuning Method}

We now consider how to train the model parameters. To expedite the tuning process, we freeze the LLM, including its embedding layer, and focus on tuning the plug-in LoRA and CIE module. Functionality speaking, the CIE module is responsible for extracting collaborative information and making it usable for LLM in recommendation, while the LoRA module assists the LLM in learning the recommendation task. To tune them, one intuitive approach is to directly train them simultaneously. However, because of the significant reliance on collaborative information, training them both from scratch concurrently may negatively impact LLM recommendations in cold scenarios. 
To address this, we propose a two-step tuning approach, tuning each component separately as follows:




\vspace{+5pt}
\noindent $\bullet$ \textbf{Step 1. Tuning the LoRA Module}.
To endow the cold-start recommendation capabilities of LLM, our initial focus is on fine-tuning the LoRA module to learn recommendation tasks independently of collaborative information. During this step, we exclude the collaborative information-related portions of the prompt, which are denoted by the content with underlines in the prompt template. Instead, we solely utilize the remaining text-only segment of the prompt to generate predictions and minimize prediction errors for tuning the LoRA module to learning recommendation. Formally, this can be expressed as:
\begin{equation}\label{eq:tuning-1}
    \hat{\Theta}^{\prime}  = argmin_{\Theta^{\prime}} \sum_{(u,i,y)\in \mathcal{D}} \ell(h_{\hat{\Theta}+\Theta^{\prime}}(E_t),y),  
\end{equation}
where $E_t$ represents the sequence of embeddings for the text-only prompt, fully obtained through the tokenization and embedding lookup in the LLM; $\ell$ denotes the recommendation loss, which is implemented as the binary cross-entropy (BCE) loss; $h_{\hat{\Theta}+\Theta^{\prime}}(E_{t})$ represents the LLM's prediction using $E_{t}$.
and $\hat{\Theta}^{\prime}$ denotes the learned parameters for the LoRA module.



\vspace{+5pt}
\noindent \textbf{$\bullet$ Step 2. Tuning the CIE Module.} 
In this step, we tune the CIE module while keeping all other components frozen. The objective of this tuning step is to enable the CIE module to learn how to extract and map collaborative information effectively for LLM usage in recommendations. To achieve this, we utilize prompts containing collaborative information, which are constructed using the full template, to generate predictions and tune the CIE model to minimize prediction errors. Formally, we solve the following optimization problem:
\begin{equation}\label{eq:tuning-2}
    min_{\Omega} \sum_{(u,i,y)\in \mathcal{D}} \ell(h_{\hat{\Theta}+\hat{\Theta}^{\prime}}(E),y), 
\end{equation}
where $E$ represents the sequence of embeddings for the full prompt, obtained through both the CIE module and the LLM's embedding lookup as shown in Equation~\eqref{eq:hybird}. $\Omega$ denotes the model parameters of the CIE module that we aim to train, for which, we consider two different choices: 
\begin{itemize} [leftmargin=*]
    \item $\Omega={\phi}$, implying that we only tune the mapping layer $g_{\phi}$ while utilizing a well-trained  collaborative model $f_{\psi=\hat{\psi}}$ in the CIE, where $\hat{\psi}$ represents pre-trained parameters for $f_{\psi}$ (with BCE).
    \item $\Omega=\{\phi,\psi\}$, meaning we train both the conventional collaborative model $f_{\psi}$ and the mapping layer $g_{\phi}$ within the CIE module.
\end{itemize}
We believe both choices are viable. The first option may be faster since it focuses solely on tuning the mapping function. However, the second option has the potential to lead to better performance as it can more seamlessly integrate collaborative information into LLM with fewer constraints from the collaborative model.

The above two steps are executed only once. It's worth noting that in step 2, we exclusively tune the CIE module without fine-tuning the LoRA to utilize collaborative information. The rationale behind this is as follows: after step 1, the LLM has already acquired the capability to perform recommendation tasks, \ie inferring the matching between users and items within the token embedding space. Collaborative information would be also leveraged based on inferring the matching. Once it is mapped into the token embedding space, we believe it can be effectively used by the LLM for recommendations without further tuning the LoRA module.


\vspace{+5pt}
\subsection{Discussion}
\textbf{Relation to Soft Prompt Tuning.} When considering our method without the LoRA module, it can be seen as a variant of soft prompt tuning in recommendation systems, with collaborative embeddings serving as the soft prompts. However, two distinct differences set our approach apart: 1) the soft prompt utilized by the LLM retains a low-rank characteristic, as it is derived from low-rank representations of conventional collaborative models; and 2) the collaborative model can provide valuable constraints and priors for learning the soft prompt, offering additional guidance regarding the collaborative information. These two factors enhance the efficacy of our method in capturing collaborative information and encoding personalized information more effectively.

\vspace{+5pt}
\noindent \textbf{Inference Efficiency.} 
We acknowledge that a significant challenge for LLMRec, including our CoLLM, is its relatively high computational cost, posing impediments to practical applications. However, a range of acceleration techniques tailored for LLMs has emerged, showcasing promising outcomes, such as caching and reusing~\cite{gim2023prompt}. LLMRec, including our CoLLM, could potentially benefit from these techniques to enhance their inference efficiency.  Given that our objective is to incorporate collaborative information for the improvement of recommendation quality, the exploration of acceleration methods is deferred to future research. Additionally, it is noteworthy that, in comparison to existing LLMRec methods (such as TALLRec), our CoLLM just introduces the CIE module. The module is much smaller when compared to LLMs, and its training does not involve the updating of the LLM. Consequently, CoLLM would not introduce excessive additional overhead in both training and inference, as demonstrated in Section~\ref{sec:exp-efficicay}.

\section{Experiments}


In this section, we perform experiments to answer two research questions: \textbf{RQ1}: Can our proposed CoLLM effectively augment LLMs with collaborative information to improve recommendation, in comparison to existing methods?  \textbf{RQ2}: What impact do our design choices have on the performance/efficiency of the proposed method? How does the method perform on other datasets and LLM backbones? 







\subsection{Experimental Settings}
We conduct our experiments on two datasets:
\begin{itemize}[leftmargin=*]
    \item \textbf{ML-1M}~\cite{movielendata} refers to the  well-known movie recommendation dataset --- MovieLens-1M\footnote{\url{https://grouplens.org/datasets/movielens/1m/}}. 
    This dataset contains user ratings on movies, collected between 2000 and 2003, with ratings on a scale of 1 to 5. We convert these ratings into binary labels using a threshold~\cite{wenjieDiffusion} of 3. Ratings greater than 3 are labeled as ``positive'' ($y=1$), while the rest are labeled as ``negative'' ($y=0$).

    \item \textbf{Amazon-Book}~\cite{amazonbook} refers to a book recommendation dataset, the ``book'' subset of the famous Amazon Product Review dataset\footnote{\url{https://cseweb.ucsd.edu/~jmcauley/datasets.html#amazon_reviews}}. It compiles user reviews of books from Amazon, collected between 1996 and 2018, with review scores ranging from 1 to 5. We transform these review scores into binary labels using a threshold\footnote{A higher value is utilized to prevent significant imbalance between positive and negative data.} of 4.
\end{itemize}

To better simulate real-world recommendation scenarios and prevent data leakage~\cite{data_Leakage_1, zhang2020retrain}, we split the dataset into training, validation, and testing sets based on the interaction timestamp. Specifically, for ML-1M, we preserve the interactions from the most recent twenty months, using the first 10 months for training, the middle 5 months for validation, and the last 5 months for testing. As for the Amazon-Book dataset, given its large scale, 
we just preserve interactions from the year 2017 (including about 4 million interactions), allocating the first 11 months for training, and the remaining two half months for validation and testing, respectively. Given the sparse nature of the Amazon-Book dataset, we filtered out users and items with fewer than 20 interactions to ensure data quality for measuring warm-start performance. The statistical information of the processed dataset is available in Table~\ref{tab:dataStatics}. 

\begin{table}[t]
\caption{Statistics of the evaluation datasets.}
\vspace{-5pt}
\label{tab:dataStatics}
\centering
\renewcommand\arraystretch{1.1}
\resizebox{0.43\textwidth}{!}{%
\begin{tabular}{cccccc}
\hline
Dataset&\#Train&\#Valid&\#Test&\#User&\#Item
\\ \hline
ML-1M&33,891&10,401&7,331&839&3,256 \\
Amazon-Book&727,468&25,747&25,747&22,967&34,154\\\hline

\end{tabular}
\vspace{-5pt}
}
\end{table}

\subsubsection{Evaluation Metrics} We employ two commonly used metrics for explicit recommendation to assess the performance of studied methods: AUC, UAUC~\cite{auc-uauc} and NDCG~\cite{ndcg}. AUC is the area under the ROC curve that quantifyes the overall prediction accuracy. UAUC is derived by first computing the AUC individually for each user over the exposed items and then averaging these results across all users. 
NDCG denotes the Normalized Discounted Cumulative Gain metrics.
AUC evaluates the overall ranking quality. UAUC and NDCG provide insights into user-level ranking quality.

\begin{table*}[]
\centering
\caption{Overall performance comparison. ``Collab.'' denotes collaborative methods. ``Rel. Imp.'' denotes the relative improvement of CoLLM compared to baselines, averaged over the AUC and UAUC metrics. For a collaborative method "X", the Rel. Imp. is computed using CoLLM-X; For LLMRec methods, it is determined by comparing them to CoLLM-MF.}
\resizebox{0.95\textwidth}{!}{
\begin{tabular}{cc|cccc|cccc}
\hline
\multicolumn{2}{c|}{Dataset}                                       & \multicolumn{4}{c|}{ML-1M}           & \multicolumn{4}{c}{Amazon-Book}      \\ \hline
\multicolumn{2}{c|}{Methods}                                       & AUC    & UAUC   & NDCG   & Rel. Imp. & AUC    & UAUC   & NDCG   & Rel. Imp. \\ \hline
\multicolumn{1}{c|}{\multirow{4}{*}{Collab.}} & MF                 & 0.6482 & 0.6361 & 0.8447 & 10.3\%    & 0.7134 & 0.5565 & 0.8194 & 12.8\%    \\
\multicolumn{1}{c|}{}                         & LightGCN           & 0.5959 & 0.6499 & 0.8564 & 13.2\%    & 0.7103 & 0.5639 & 0.8245 & 11.0\%    \\
\multicolumn{1}{c|}{}                         & SASRec             & 0.7078 & 0.6884 & 0.8612 & 1.9\%     & 0.6887 & 0.5714 & 0.8244 & 8.4\%     \\
\multicolumn{1}{c|}{}                         & DIN                & 0.7166 & 0.6459 & 0.8496 & 4.9\%     & 0.8163 & 0.6145 & 0.8419 & 3.2\%     \\ \hline
\multicolumn{1}{c|}{LM+Collab.}               & CTRL (DIN)         & 0.7159 & 0.6492 & 0.8559 & 4.6\%     & 0.8202 & 0.5996 & 0.8363 & 4.2\%     \\ \hline
\multicolumn{1}{c|}{\multirow{3}{*}{LLMRec}}  & ICL                & 0.5320 & 0.5268 & 0.8114 & 33.8\%    & 0.4820 & 0.4856 & 0.7917 & 48.2\%    \\
\multicolumn{1}{c|}{}                         & Prompt4NR (Vicuna) & 0.7071 & 0.6739 & 0.8663 & 2.7\%     & 0.7224 & 0.5881 & 0.8346 & 10.4\%    \\
\multicolumn{1}{c|}{}                         & TALLRec            & 0.7097 & 0.6818 & 0.8711 & 1.8\%     & 0.7375 & 0.5983 & 0.8361 & 8.2\%     \\ \hline
\multicolumn{1}{c|}{\multirow{4}{*}{Ours}}    & CoLLM-MF           & 0.7295 & 0.6875 & 0.8714 & -         & 0.8109 & 0.6225 & 0.8457 & -         \\
\multicolumn{1}{c|}{}                         & CoLLM-LightGCN     & 0.7100 & 0.6967 & 0.8740 & -         & 0.8026 & 0.6149 & 0.8411 & -         \\
\multicolumn{1}{c|}{}                         & CoLLM-SASRec       & 0.7235 & 0.6990 & 0.8765 & -         & 0.7746 & 0.5962 & 0.8361 & -         \\
\multicolumn{1}{c|}{}                         & CoLLM-DIN          & 0.7353 & 0.6923 & 0.8735 & -         & 0.8245 & 0.6474 & 0.8550 & -         \\ \hline
\end{tabular}
}
\label{mainexp}
\end{table*}

\subsubsection{Compared Methods}
To fully evaluate the proposed method CoLLM, we compare it with three types of methods: conventional collaborative methods, combining language model and collaborative model methods, and LLMRec methods. Specifically, we select the following methods as baselines:

\begin{itemize}[leftmargin=*]
    \item \textbf{MF~\cite{mf}}: This refers to Matrix Factorization, one representative latent factor-based collaborative filtering method.  
     \item \textbf{LightGCN~\cite{lightgcn}}: This is a representative graph-based collaborative filtering method, which utilizes a simplified graph convolutional neural network to enhance user interest modeling.
     
      \item \textbf{SASRec~\cite{SASRec}}: This is a representative sequential recommendation method, which uses the self-attention network to encode sequential patterns for modeling user interest. It could be thought of as a collaborative method considering sequential information.

      \item \textbf{DIN~\cite{DIN}}: This is a representative collaborative CTR model, which employs an attention mechanism to activate the most relevant user behavior for adaptively learning user interest with respect to a certain item.

      \item \textbf{CTRL (DIN)~\cite{ctrl}}: This is a state-of-the-art (SOTA) method for combining language and collaborative models through knowledge distillation. We 
      utilize a DIN~\cite{DIN} as the collaborative model.
      
     \item \textbf{ICL~\cite{uncoveringGPTRec}}: This is a LLMRec method based on the In-Context Learning ability of LLM. It directly queries the original LLM for recommendations using prompts.

     \item \textbf{Prompt4NR (Vicuna)~\cite{prompt4nr}}: This is a SOTA method that \zjz{uses both fixed and soft prompts to utilize traditional Language Models (LM), such as BERT~\cite{bert}, for recommendation purposes}. We extend this method to the LLM Vicuna-7B for a fair comparison and tune the LLM with LoRA to manage computational costs.    

     \item \textbf{TALLRec~\cite{tallrec}}: This is a state-of-the-art LLMRec method that aligns LLM with recommendations through instruction tuning. We implement it on Vicuna-7B.
     
\end{itemize}


Apart from CTRL, there are other language model-based recommender models such as P5~\cite{p5} and CTR-BERT~\cite{ctr_bert}. However, these models have demonstrated weaker performance when compared to CTRL~\cite{ctrl}. Therefore, we have chosen not to include them in our comparative analysis.
Regarding BIGRec~\cite{bigrec}, it is not suitable for comparison in our setting, as it does not optimize prediction accuracy and would yield poor performance in our setting\footnote{For instance, BIGRec's highest AUC on ML-1M is only 0.56, whereas TALLRec surpasses 0.70.}
Regarding our own methods, we have implemented them across all four types of collaborative models, denoted as CoLLM-MF, CoLLM-LightGCN, CoLLM-SASRec, and CoLLM-DIN, respectively. Specifically, for collaborative models (SASRec and DIN) that incorporate historical sequences, we incorporate their sequence representation as a part of the user representation within our CIE module ($\bm{u}$ in Equation~\eqref{eq:CIE}).

\subsubsection{Implementation Details}
We implement all the compared methods using PyTorch 2.0. When not specified by the original paper, we employ Binary Cross-Entropy (BCE) as the optimization loss for all methods. For (large) language models, we use the AdamW optimizer, and for other methods, we use the Adam optimizer~\cite{adam}. 
Regarding hyperparameter tuning, we explore the learning rate within the range of [1e-2, 1e-3, 1e-4] for all methods, and the (recommendation) embedding size within the range of [64, 128, 256].
Regarding weight decay, we set it to 1e-3 for all LLM-based methods, while we tune it in the range of [1e-2, 1e-3, $\dots$, 1e-7] for all other smaller models. 
For SASRec, we establish the maximum length of historical interaction sequences in accordance with the average user interaction count in the training data, as specified in the original paper.  We adopt TALLRec's~\cite{tallrec} practice of setting the maximum sequence length to 10 for all other methods. For DIN and CTRL (DIN), we conduct additional tuning for the dropout ratio within the range of [0.2, 0.5, 0.8]. We also adjust their hidden layer size, which varies between [200$\times$80$\times$1] and [256$\times$128$\times$64$\times$1], corresponding to the two settings described in the CTRL and DIN papers. Regarding other specific parameters of the baseline models, we adhere to the configurations outlined in their original papers. For CoLLM, we set the hidden layer size of the MLP in the CIE module as ten times larger than the input size. \revision{For the LoRA module, we follow the same configuration as in the TALLRec paper, setting $r$, $alpha$, $dropout$, and $target\_modules$,  to 8, 16, 0.05, and ``[q\_proj, v\_proj]", respectively. 
}

\begin{table*}[h]
\caption{Ensemble results on the AUC metric.}
\vspace{-5pt}
\centering
\resizebox{0.65\textwidth}{!}{%
\begin{tabular}{c|ccc|cc}
\hline
            & \multicolumn{3}{c|}{Single}     & \multicolumn{2}{c}{Ensemble} \\ \hline
Methods     & MF     & TALLRec & CoLLM-MF        & MF+TALLRec & MF+CoLLM-MF        \\ \hline
ML-1M       & 0.6482 & 0.7097  & {\ul 0.7295} & 0.7239     & \textbf{0.7364} \\
Amazon-book & 0.7134 & 0.7375  & {\ul 0.8109} & 0.7782     & \textbf{0.8112} \\ \hline
\end{tabular}
}
\label{tab:ensemble}
\end{table*}

\begin{figure*}[htbp] 
\centering
\subfigure{
\includegraphics[width=2.7in, height=1.5in]{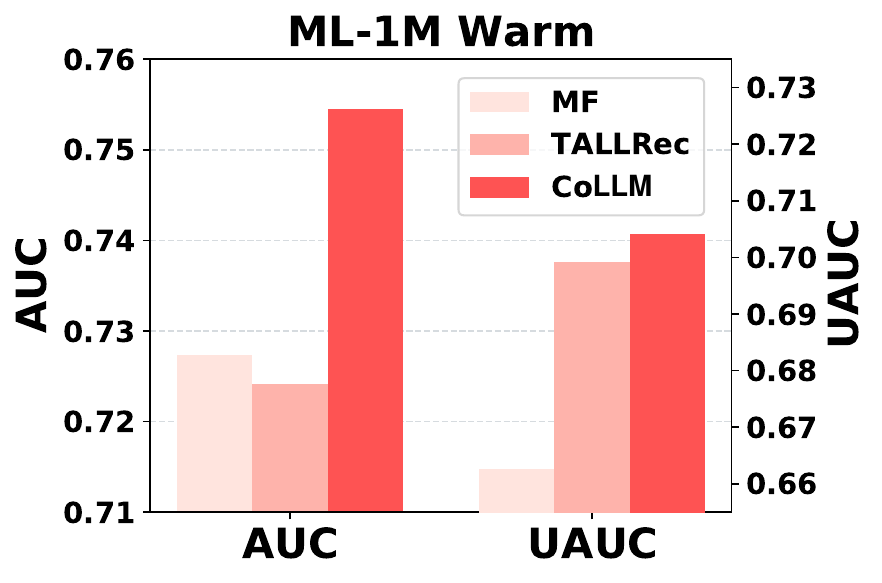} 
}
\subfigure{
\includegraphics[width=2.7in, height=1.5in]{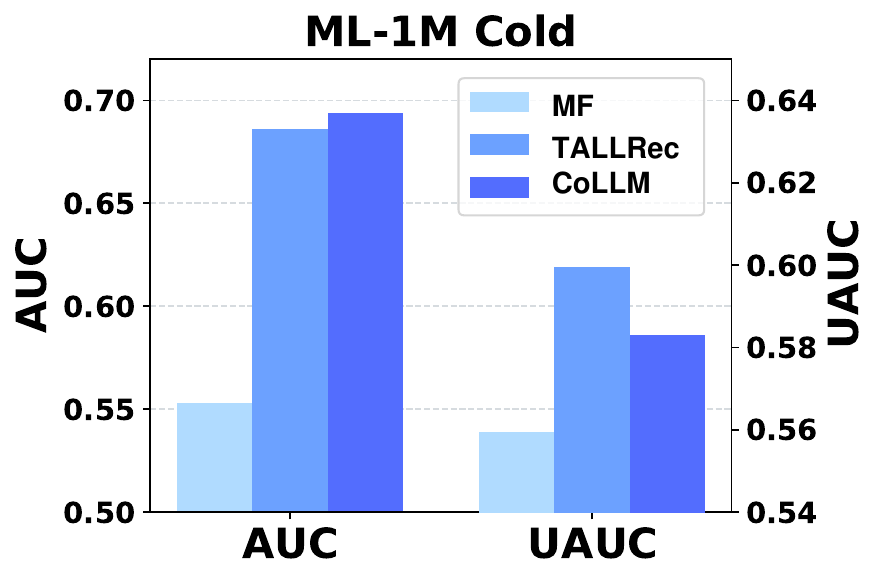}
}

\subfigure{
\includegraphics[width=2.7in, height=1.5in]{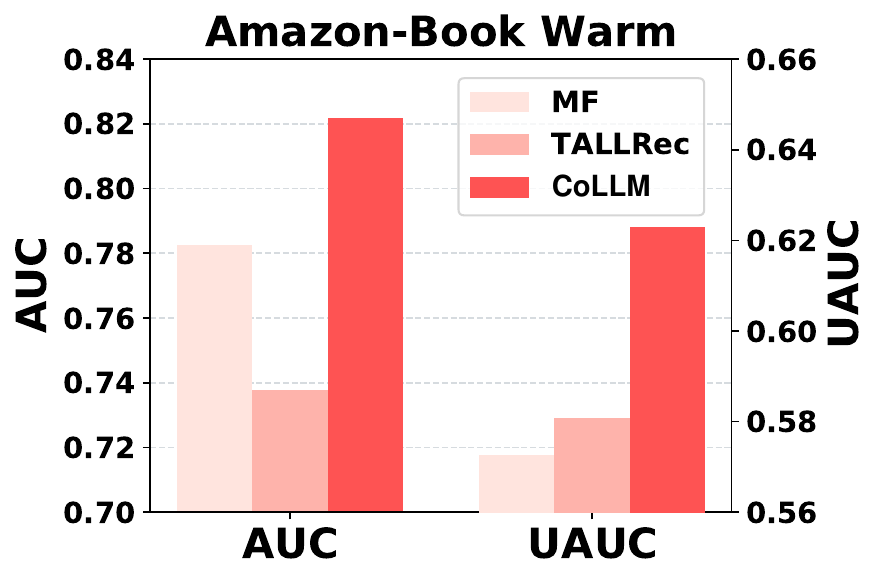}
}
\subfigure{
\includegraphics[width=2.7in, height=1.5in]{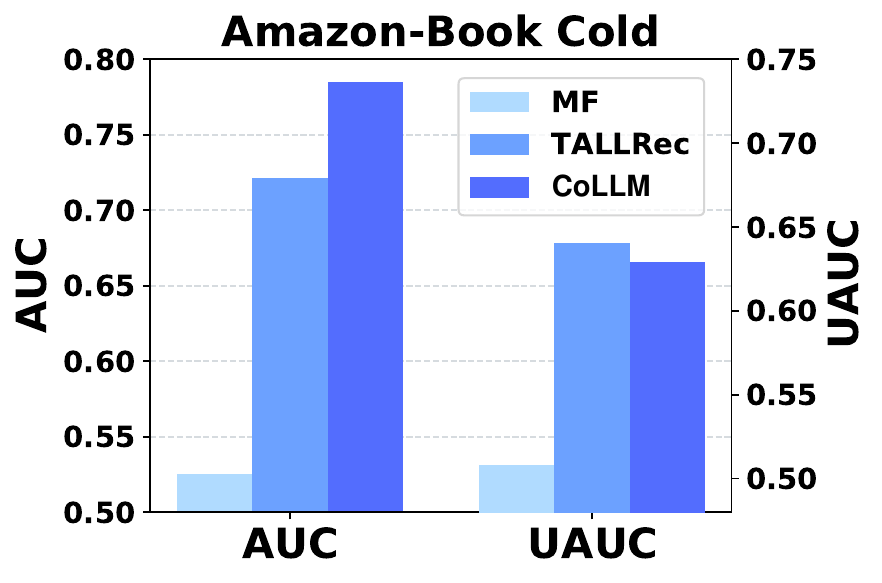}
}
\vspace{-8pt}
\caption{
Performance comparison in warm (left) and cold (right) scenarios on ML-1M and Amazon-Book.
}
\label{fig:cold-warm}
\end{figure*}

\subsection{Performance Comparison (RQ1)}
In this section, we study the recommendation performance of CoLLM over all users as
well as in different subgroups.

\subsubsection{Overall Performance} The overall performance comparison between CoLLM and baselines is summarized
in Table~\ref{mainexp} ({where the CoLLM-DIN results on ML-1M are obtained by tuning the whole CIE}). From the table, we have the following observations:
\begin{itemize}[leftmargin=*]
    \item When compared to baselines, \zy{the best version of CoLLM outperforms them in both metrics on the two datasets.} The results confirm the excellence of our approach.
    
    \item In comparison between the best LLMRec baseline (TALLRec) and collaborative models, it outperforms MF, LightGCN, and SASRec but falls short of beating DIN. However, after introducing collaborative information into LLMRec using CoLLM, LLM consistently achieves performance improvements (except for UAUC/NDCG for SASRec on Amazon-Book) and surpasses the corresponding collaborative baselines. This demonstrates the necessity of incorporating collaborative information.

    \item When focusing on LLMRec methods, the ICL method consistently produces the weakest results, in line with previous findings~\cite{tallrec}. This underscores LLM's inherent limitation in recommendation and emphasizes the importance of tuning LLM for recommendation tasks. 
    Interestingly, when we scrutinize Prompt4NR (Vicuna), it not only fine-tunes LLM itself but also incorporates some adaptable prompts, bearing resemblances to CoLLM. Nevertheless, it not only falls short of matching CoLLM's performance but even lags behind TALLRec, which exclusively fine-tunes LLM. This suggests that the improvements in CoLLM stem from its collaborative information modeling mechanism rather than only relying on adaptively updatable prompts.

    \item  Regarding the CTRL approach, which integrates LM and collaborative models, it exhibits the capability to enhance one metric while adversely affecting another across both datasets. 
    This implies its capability to effectively harness both the LM's strengths and collaborative information for enhanced recommendations is limited.
    The operation of CTRL involves initially distilling LM information into DIN and subsequently fine-tuning DIN to incorporate collaborative information. This operation faces an inherent challenge in terms of the forgetting problem, resulting in the loss of language model information. Additionally, it continues to depend on DIN for recommendations, lacking the utilization of the LM model's inherent capabilities, such as reasoning. Our CoLLM aligns collaborative information with LLMs while still relying on the LLM for predictions, effectively mitigating these limitations and consistently improving performance.

    \item When implementing CoLLM's CIE module with various collaborative models, it consistently yields improvements compared to both the respective collaborative method baselines and LLMRec baselines in almost all cases. This showcases CoLLM's flexibility in incorporating different collaborative modeling mechanisms. Furthermore, it's worth highlighting that CoLLM's performance is roughly positively correlated with the performance of the corresponding collaborative model. This suggests that introducing better collaborative modeling mechanisms could contribute to enhancing the performance of CoLLM.
    

\end{itemize}





\vspace{+5pt}
\noindent\textbf{Ensemble.}
We have not included the ensemble method as our baseline above, as it could also be applied to our approach. Here, we conduct a detailed study.
To do so, we employ ensemble averaging~\cite{ensemble} to combine the MF and TALLRec models, and then compare the results with our CoLLM-MF approach. Furthermore, we investigate whether ensembling CoLLM-MF with MF can achieve further improvements through ensemble averaging. The performance of these methodologies is summarized in Table~\ref{tab:ensemble}.
The table indicates that combining MF and TALLRec via ensemble averaging consistently yields inferior results compared to CoLLM-MF. However, applying ensemble averaging to CoLLM-MF still leads to some marginal improvements. These findings underscore the superiority of CoLLM's mechanism for integrating collaborative information with LLMs, surpassing mere ensemble techniques and facilitating the full utilization of LLMs' capabilities.

\subsubsection{Performance in Warm and Cold Scenarios} 
Previous research has demonstrated that LLMRec excels in cold-start scenarios~\cite{tallrec}, while collaborative information is advantageous for modeling user interests when rich data is available~\cite{rella, 9429954}. CoLLM seeks to incorporate collaborative information into LLMRec, aiming to make it perform well in both warm and cold scenarios. To assess the success of this goal, we conduct a detailed examination of the methods' performance in warm and cold scenarios. In particular, we divide the testing set into warm and cold subsets: the warm subset comprises interactions between users and items that have appeared at least three times in the training dataset, while the cold subset includes the remaining interactions. 
\revision{Notably, our cold-start scenario is not strictly cold, as it allows users/items to have a few interactions.}
Without losing generality, we primarily compare MF, TALLRec, and CoLLM in terms of their performance using these two subsets. Our findings are presented in Figure~\ref{fig:cold-warm}, from which we make three observations:
\begin{itemize}[leftmargin=*]
   \item In the warm scenario, across two datasets, TALLRec exhibits a lower AUC score compared to MF, and MF in turn is outperformed by CoLLM. In terms of UAUC, MF falls short of TALLRec, which lags behind CoLLM. These results suggest that, at the very least, in the overall evaluation (AUC) context, existing LLMRec (TALLRec) has shortcomings in warm scenarios. However, the introduction of collaborative information can lead to improvements in warm scenarios. 
   \item 
   In the cold scenario, both TALLRec and CoLLM clearly outperform MF. This highlights the advantages of LLMRec methods in cold scenarios, while collaborative methods lack the proficiency to handle cold scenarios. 
   Meanwhile, CoLLM broadly maintains comparability with TALLRec, performing better in AUC and slightly worse in UAUC.  These imply that CoLLM can still effectively leverage the strengths of LLMRec in cold scenarios.
\end{itemize}
Overall, CoLLM has demonstrated remarkable improvements over TALLRec in warm scenarios while maintaining its proficiency in cold scenarios. This underscores CoLLM's successful integration of collaborative information to achieve the goal of enabling LLMRec to perform effectively in both cold and warm scenarios.

\begin{table}[t]
\centering
\caption{
Results of the ablation studies over CoLLM with respect to the CIE module.
}
\label{tab:my-table}
\centering
\resizebox{0.42\textwidth}{!}{%
\begin{tabular}{c|cc|cc}
\hline
Dataset       & \multicolumn{2}{c|}{ML-1M} & \multicolumn{2}{c}{Amazon-Book} \\ \hline
Methods       & AUC          & UAUC        & AUC            & UAUC           \\ \hline
CoLLM-MF & 0.7295       & 0.6875      & 0.8133         & 0.6314         \\ \hline
w/o CIE       & 0.7097       & 0.6818      & 0.7375         & 0.5983         \\ \hline
w/ UI-token   & 0.7214       & 0.6563      & 0.7273         & 0.5956         \\ \hline
\end{tabular}%
}
\label{tab:ab-model}
\end{table}

\subsection{In-depth Analysis (RQ2)}
In this subsection, we conduct experiments to study the influence of different design choices on the CoLLM.


\subsubsection{The Effect of CIE Module} 
We begin by investigating the impact of model architecture designs. The central element of our designs is the introduction of the CIE module to extract collaborative information for LLMs. To assess its influence on CoLLM's performance, we compare CoLLM-MF with two variants: 1) the variant that directly omits the CIE module (referred to as ``w/o CIE''); and 2) the variant that excludes the CIE module but instead directly introduces tokens and token embeddings to represent users and items in the LLM (referred to as ``w/ UI-token''). The first variant is equal to TALLRec. The second variant follows the straightforward approach for modeling collaborative information described in Section~\ref{sec:into}.  The comparison results are summarized in Table~\ref{tab:ab-model}. 

According to the figure, the ``w/o CIE'' variant falls short of the original CoLLM. 
\revision{This result underscores the core role of the CIE module in CoLLM to enhance the performance.}
The ``w/ UI-token'' variant also yields inferior performance compared to CoLLM and even performs worse than the variant lacking collaborative information modeling (\ie ``w/o CIE''). This observation confirms that directly introducing tokens for users and items in the LLM cannot effectively capture collaborative information for it. The rationale behind this could be that incorporating tokens (along with token embeddings) for encoding collaborative information increases tokenization redundancy within the LLM and subsequently diminishes the model's compression efficiency, leading to a reduction in predictive capabilities, as discussed in~\cite{LLMisCompression}. In contrast, our method employs a traditional collaborative model for encoding, effectively maintaining the modeled collaborative information in a low-rank state to reduce redundancy.     


\begin{figure*}[t]
\centering  
\subfigure{
\includegraphics[width=2.75in, height=1.55in]{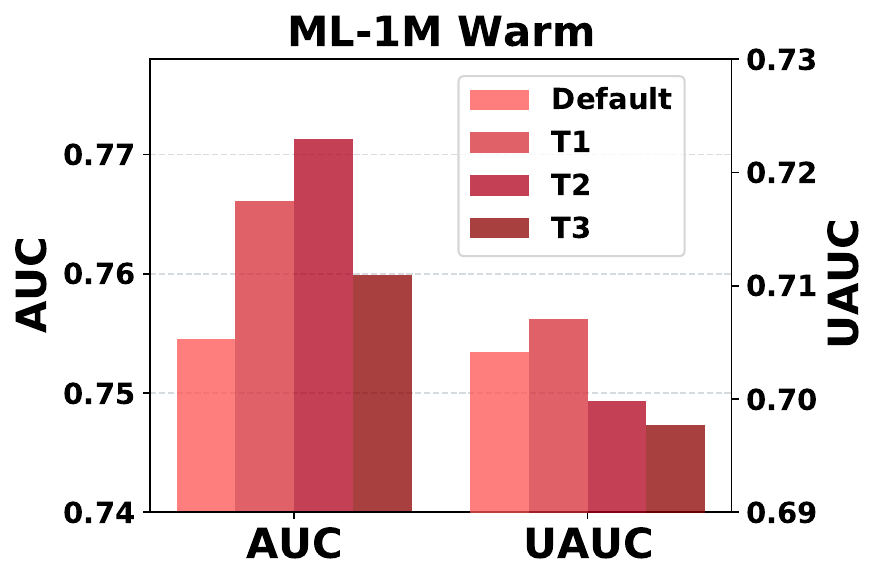}}
\subfigure{
\includegraphics[width=2.75in, height=1.55in]{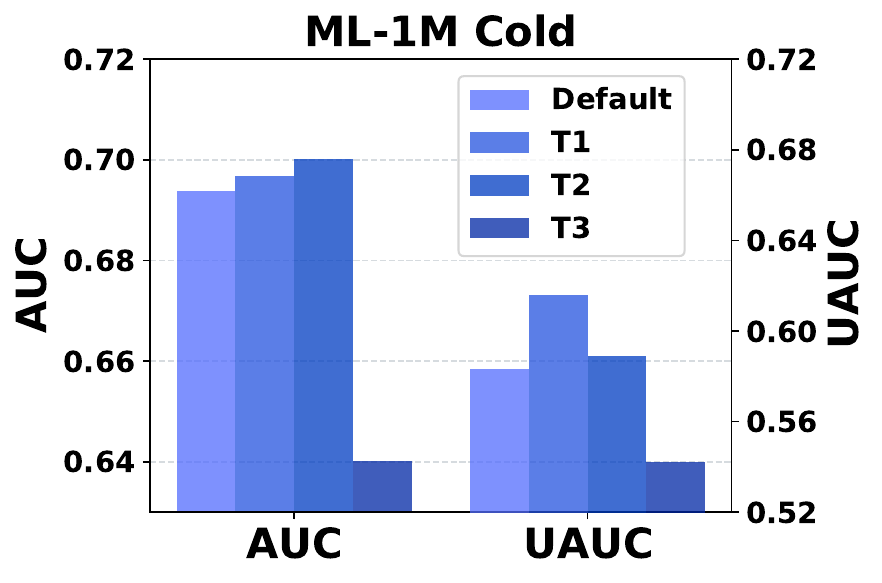}}
\vspace{-10pt}
\caption{Performance of different tuning strategies for CoLLM on ML-1M in warm and cold scenarios.}
\label{fig:ab-tuning-domain}
\end{figure*}


\begin{table}[t]
\centering
\caption{Overall performance of CoLLM with different tuning strategies.}

\label{tab:ab-tuning-overall}
\resizebox{0.42\textwidth}{!}{%
\begin{tabular}{c|cc|cc}
\hline
Dataset        & \multicolumn{2}{c|}{ML-1M} & \multicolumn{2}{c}{Amazon-Book} \\ \hline
Tuning Methods & AUC          & UAUC        & AUC            & UAUC           \\ \hline
Default        & 0.7295       & 0.6875      & 0.8109         & 0.6225         \\ \hline
T1             & 0.7360       & 0.6946      & 0.8154         & 0.6139         \\ \hline
T2             & 0.7418       & 0.6906      & 0.8288         & 0.6352         \\ \hline
T3             & 0.7131       & 0.6661      & 0.8104         & 0.5753         \\ \hline
\end{tabular}%
}
\end{table}

\subsubsection{The Influence of Tuning Choices} 
We now delve into the impact of training choices on CoLLM's performance. Our default approach involves a two-step tuning strategy. Initially, we exclusively take textual information to train the LoRA module for recommendation task learning. Subsequently, we tune the mapping layer of the CIE module while keeping the collaborative model fixed as pre-trained (\ie tuning $\Omega=\phi$ while retaining $\psi=\hat{\psi}$ after step 1). In this subsection, we further explore the following tuning strategies: 
\begin{itemize}[leftmargin=*]
    \item T1, aligning with our default two-step tuning but tunes the entire CIE model ($\Omega=\{\phi,\psi\}$) with the collaborative model ($\psi$) initialized to the pre-trained one ($\hat{\psi}$) in the second step.
    \item T2, following the default two-step tuning approach but tunes the entire CIE model ($\Omega=\{\phi,\psi\}$) from scratch in the second step.
    \item T3, employing a one-step tuning approach, directly tuning the LoRA module and the CIE's mapping layer simultaneously while fixing the pre-trained collaborative model.
\end{itemize}
We compare these methods in terms of their overall performance, as shown in Table~\ref{tab:ab-tuning-overall}, as well as their performance in warm and cold scenarios, as indicated in Figure~\ref{fig:ab-tuning-domain}. \zy{Please note that in the figure, we have omitted the results on the Amazon-book, as they exhibit similar phenomena to those on the ML-1M, to save space.}


Based on the figure and table, we observe that within our two-step update framework, the additional tuning of the collaborative model in CIE (\ie T1 and T2) can yield additional performance improvements in most cases, as expected. The additional tuning allows the captured collaborative information to be seamlessly integrated by LLMs. However, it does introduce extra computational overhead and slower convergence rates, \eg requiring at least five times the training effort in ML-1M. As a result, we opt for more efficient approaches.
Furthermore, when comparing the single-step tuning method, T3, with the other two-step methods, we notice that it usually exhibits relatively inferior results, particularly in cold scenarios, where its performance decline is quite noticeable. This underscores the significance of the first step in our two-step training, which uses text-only data to learn recommendation tasks, to ensure recommendation performance in cold start scenarios.


\subsubsection{Efficiency}\label{sec:exp-efficicay} 
\zy{Efficiency challenges pose a significant impediment to the application of the LLMRec. We next investigate how our design, which incorporates collaborative information, influences the training and inference efficiency of LLMRec. In terms of training, the primary additional cost in our approach arises from the training of the CIE module in our two-step tuning. Fortunately, the pre-training of the collaborative module (\ie $f_{\psi=\hat{\psi}}$) in the CIE alleviates the need for extensive additional training in CoLLM when compared to the representative LLMRec method TALLRec. Table~\ref{tab:time-cost} illustrates that, under identical resource conditions\footnote{Training utilizes two Nvidia A100 80G GPUs, while testing is performed on a single GPU of the same type.},
the training time of CoLLM increases by only approximately 15.5\%, averaged across the two datasets, compared to the baseline TALLRec. }

\zy{Concerning inference, the cost associated with the CIE module is anticipated to be negligible due to its significantly smaller scale compared to the LLM. The primary additional cost in our method arises from the extra tokens required to describe collaborative information in our prompt template (as indicated in the underlined part) in Section~\ref{sec:prompt-template}. These additional tokens constitute only a small proportion of the total prompts. Consequently, the supplementary inference cost is expected to be relatively modest. As demonstrated in Table~\ref{tab:time-cost}, CoLLM incurs only a 12.5\% increase in total inference cost on average across the two datasets.}

\begin{table}[t]
\centering
\caption{Training and Total Inference Time Comparison: TALLRec vs. CoLLM-MF. "$\Delta$" indicates the relative cost improvement of CoLLM-MF over TALLRec. "Book" is short for "Amazon-book".}
\label{tab:time-cost}
\resizebox{0.45\textwidth}{!}{%
\begin{tabular}{c|cc|cc}
\hline
 & \multicolumn{2}{c|}{Train Time} & \multicolumn{2}{c}{Inference Time} \\ \hline
Dataset & ML-1M & Book & ML-1M & Book \\ \hline
TALLRec & 32min &  354min & 72s & 360s \\
CoLLM-MF &  36min &  418min & 82s & 398s \\
$\Delta$ & 13\% & 18\% & 14\% & 11\% \\ \hline
\end{tabular}%
}
\end{table}

\subsubsection{Method Generalization}
This section investigates whether our method can effectively apply to other datasets and LLM backbones.

\begin{table}
\caption{Performance comparison on Qwen2-1.5B backbone across ML-1M and Amazon-Book datasets.}
\vspace{-5pt}
\centering
\resizebox{0.45\textwidth}{!}{%
\begin{tabular}{l|ll|ll}
\hline
Dataset  & \multicolumn{2}{l|}{ML-1M} & \multicolumn{2}{l}{Amazon-Book} \\ \hline
Metric   & AUC          & UAUC        & AUC            & UAUC           \\ \hline
MF       & 0.6482       & 0.6361      & 0.7134         & 0.5565         \\
TALLRec  & 0.7027       & 0.6638      & 0.7256         & 0.5830         \\
CoLLM-MF & 0.7354       & 0.6950      & 0.8068         & 0.6147         \\ \hline
\end{tabular}
}
\label{tab:qwen2}
\end{table}

\vspace{+5pt}
\noindent\textbf{Other LLM Backbone.} 
\revision{To further validate the effectiveness of our approach on different LLM backbones, we implement CoLLM-MF and the top-performing LLMRec baseline, TALLRec, using the Qwen2-1.5B~\cite{yang2024qwen2} backbone to compare their performance. The results, summarized in Table~\ref{tab:qwen2}, show that CoLLM consistently outperforms both TALLRec and MF. This demonstrates the general applicability of our method across different LLM backbones.
}

\vspace{+5pt}
\noindent\textbf{Other Dataset.}
\revision{
To further validate the effectiveness of our approach across different datasets, we have included two additional datasets: Video Games and CDs \& Vinyl from the Amazon dataset. We take CoLLM-MF as the study example and compare it with the most related baselines MF and TALLRec. The results demonstrate that our method could still effectively enhance LLMRec by incorporating collaborative information. Specifically, on the Video Games dataset, the AUC values for MF, TALLRec, and CoLLM-MF are 0.6161, 0.7356, and 0.7440, respectively. Similarly, on the CDs \& Vinyl dataset, the AUC values are 0.6957, 0.6607, and 0.7237, respectively. The consistent superior performance of the proposed method across different datasets indicates its general effectiveness.}

\section{Conclusion}
In this study, we underscore the significance of collaborative information modeling in enhancing recommendation performance for LLMRec, particularly in warm scenarios. We introduce CoLLM, a novel approach tailored to incorporating collaborative information for LLMRec.
By externalizing traditional collaborative models for LLMs, CoLLM not only ensures effective collaborative information modeling but also provides flexibility in adjusting the modeling mechanism. Extensive experimental results illustrate the effectiveness and adaptability of CoLLM, successfully enabling LLM to excel in both warm and cold recommendation scenarios. Currently, our experiments have been exclusively conducted on Vicuna-7B. In the future, we will explore other LLMs. Moreover, considering the evolving nature of collaborative information in the actual world, we intend to investigate CoLLM's incremental learning capabilities.

\small
\bibliographystyle{IEEEtran}
\bibliography{reference}

\begin{thebibliography}{10}
\providecommand{\url}[1]{#1}
\csname url@samestyle\endcsname
\providecommand{\newblock}{\relax}
\providecommand{\bibinfo}[2]{#2}
\providecommand{\BIBentrySTDinterwordspacing}{\spaceskip=0pt\relax}
\providecommand{\BIBentryALTinterwordstretchfactor}{4}
\providecommand{\BIBentryALTinterwordspacing}{\spaceskip=\fontdimen2\font plus
\BIBentryALTinterwordstretchfactor\fontdimen3\font minus \fontdimen4\font\relax}
\providecommand{\BIBforeignlanguage}[2]{{%
\expandafter\ifx\csname l@#1\endcsname\relax
\typeout{** WARNING: IEEEtran.bst: No hyphenation pattern has been}%
\typeout{** loaded for the language `#1'. Using the pattern for}%
\typeout{** the default language instead.}%
\else
\language=\csname l@#1\endcsname
\fi
#2}}
\providecommand{\BIBdecl}{\relax}
\BIBdecl

\bibitem{instructGPT}
L.~Ouyang, J.~Wu, X.~Jiang, D.~Almeida, C.~Wainwright, P.~Mishkin, C.~Zhang, S.~Agarwal, K.~Slama, A.~Ray \emph{et~al.}, ``Training language models to follow instructions with human feedback,'' \emph{Advances in Neural Information Processing Systems}, pp. 27\,730--27\,744, 2022.

\bibitem{GPT3}
T.~Brown, B.~Mann, N.~Ryder, M.~Subbiah, J.~D. Kaplan, P.~Dhariwal, A.~Neelakantan, P.~Shyam, G.~Sastry, A.~Askell \emph{et~al.}, ``Language models are few-shot learners,'' \emph{Advances in neural information processing systems}, pp. 1877--1901, 2020.

\bibitem{llama}
H.~Touvron, T.~Lavril, G.~Izacard, X.~Martinet, M.-A. Lachaux, T.~Lacroix, B.~Rozi{\`e}re, N.~Goyal, E.~Hambro, F.~Azhar \emph{et~al.}, ``Llama: Open and efficient foundation language models,'' \emph{arXiv:2302.13971}, 2023.

\bibitem{zhao-survey}
W.~X. Zhao, K.~Zhou, J.~Li, T.~Tang, X.~Wang, Y.~Hou, Y.~Min, B.~Zhang, J.~Zhang, Z.~Dong \emph{et~al.}, ``A survey of large language models,'' \emph{arXiv:2303.18223}, 2023.

\bibitem{blip2}
J.~Li, D.~Li, S.~Savarese, and S.~C.~H. Hoi, ``Blip-2: Bootstrapping language-image pre-training with frozen image encoders and large language models,'' in \emph{International Conference on Machine Learning}, 2023, pp. 19\,730--19\,742.

\bibitem{singhal2023large}
K.~Singhal, S.~Azizi, T.~Tu, S.~S. Mahdavi, J.~Wei, H.~W. Chung, N.~Scales, A.~Tanwani, H.~Cole-Lewis, S.~Pfohl \emph{et~al.}, ``Large language models encode clinical knowledge,'' \emph{Nature}, vol. 620, no. 7972, pp. 172--180, 2023.

\bibitem{singh2023progprompt}
I.~Singh, V.~Blukis, A.~Mousavian, A.~Goyal, D.~Xu, J.~Tremblay, D.~Fox, J.~Thomason, and A.~Garg, ``Progprompt: Generating situated robot task plans using large language models,'' in \emph{2023 IEEE International Conference on Robotics and Automation (ICRA)}, 2023, pp. 11\,523--11\,530.

\bibitem{10417790}
L.~Yang, H.~Chen, Z.~Li, X.~Ding, and X.~Wu, ``Give us the facts: Enhancing large language models with knowledge graphs for fact-aware language modeling,'' \emph{IEEE Transactions on Knowledge and Data Engineering}, pp. 1--20, 2024.

\bibitem{LLMrec-cold}
S.~Sanner, K.~Balog, F.~Radlinski, B.~Wedin, and L.~Dixon, ``Large language models are competitive near cold-start recommenders for language-and item-based preferences,'' in \emph{Proceedings of the 17th ACM Conference on Recommender Systems}, 2023, pp. 890--896.

\bibitem{ehcheng-survey}
L.~Wu, Z.~Zheng, Z.~Qiu, H.~Wang, H.~Gu, T.~Shen, C.~Qin, C.~Zhu, H.~Zhu, Q.~Liu \emph{et~al.}, ``A survey on large language models for recommendation,'' \emph{arXiv:2305.19860}, 2023.

\bibitem{qing-llmrec}
Z.~Zhao, W.~Fan, J.~Li, Y.~Liu, X.~Mei, Y.~Wang, Z.~Wen, F.~Wang, X.~Zhao, J.~Tang, and Q.~Li, ``Recommender systems in the era of large language models (llms),'' \emph{{IEEE} Trans. Knowl. Data Eng.}, pp. 1--20, 2024.

\bibitem{tallrec}
K.~Bao, J.~Zhang, Y.~Zhang, W.~Wang, F.~Feng, and X.~He, ``Tallrec: An effective and efficient tuning framework to align large language model with recommendation,'' in \emph{Proceedings of the 17th ACM Conference on Recommender Systems}, 2023, p. 1007–1014.

\bibitem{zhang2023chatgpt}
J.~Zhang, K.~Bao, Y.~Zhang, W.~Wang, F.~Feng, and X.~He, ``Is chatgpt fair for recommendation? evaluating fairness in large language model recommendation,'' in \emph{Proceedings of the 17th ACM Conference on Recommender Systems}, 2023, pp. 993--999.

\bibitem{mf}
Y.~Koren, R.~Bell, and C.~Volinsky, ``Matrix factorization techniques for recommender systems,'' \emph{Computer}, vol.~42, no.~8, pp. 30--37, 2009.

\bibitem{amazonbook}
R.~He and J.~McAuley, ``Ups and downs: Modeling the visual evolution of fashion trends with one-class collaborative filtering,'' in \emph{proceedings of the 25th international conference on world wide web}, 2016, pp. 507--517.

\bibitem{LMRecSys}
Y.~Zhang, H.~DING, Z.~Shui, Y.~Ma, J.~Zou, A.~Deoras, and H.~Wang, ``Language models as recommender systems: Evaluations and limitations,'' in \emph{I (Still) Can't Believe It's Not Better! NeurIPS 2021 Workshop}, 2021.

\bibitem{uncoveringGPTRec}
S.~Dai, N.~Shao, H.~Zhao, W.~Yu, Z.~Si, C.~Xu, Z.~Sun, X.~Zhang, and J.~Xu, ``Uncovering chatgpt’s capabilities in recommender systems,'' in \emph{Proceedings of the 17th ACM Conference on Recommender Systems}, 2023, p. 1126–1132.

\bibitem{chat-rec}
Y.~Gao, T.~Sheng, Y.~Xiang, Y.~Xiong, H.~Wang, and J.~Zhang, ``Chat-rec: Towards interactive and explainable llms-augmented recommender system,'' \emph{arXiv}, 2023.

\bibitem{llmsKnowPref}
W.-C. Kang, J.~Ni, N.~Mehta, M.~Sathiamoorthy, L.~Hong, E.~Chi, and D.~Z. Cheng, ``Do llms understand user preferences? evaluating llms on user rating prediction,'' \emph{arXiv}, 2023.

\bibitem{bigrec}
K.~Bao, J.~Zhang, W.~Wang, Y.~Zhang, Z.~Yang, Y.~Luo, F.~Feng, X.~He, and Q.~Tian, ``A bi-step grounding paradigm for large language models in recommendation systems,'' \emph{arXiv:2308.08434}, 2023.

\bibitem{instructfollows}
J.~Zhang, R.~Xie, Y.~Hou, W.~X. Zhao, L.~Lin, and J.-R. Wen, ``Recommendation as instruction following: A large language model empowered recommendation approach,'' \emph{arXiv:2305.07001}, 2023.

\bibitem{rella}
J.~Lin, R.~Shan, C.~Zhu, K.~Du, B.~Chen, S.~Quan, R.~Tang, Y.~Yu, and W.~Zhang, ``Rella: Retrieval-enhanced large language models for lifelong sequential behavior comprehension in recommendation,'' in \emph{The Web Conference 2024}.

\bibitem{ctrl}
X.~Li, B.~Chen, L.~Hou, and R.~Tang, ``Ctrl: Connect tabular and language model for ctr prediction,'' \emph{arXiv}, 2023.

\bibitem{lewu-survey}
L.~Wu, X.~He, X.~Wang, K.~Zhang, and M.~Wang, ``A survey on accuracy-oriented neural recommendation: From collaborative filtering to information-rich recommendation,'' \emph{{IEEE} Trans. Knowl. Data Eng.}, vol.~35, pp. 4425--4445, 2023.

\bibitem{9601264}
X.~Luo, Y.~Zhou, Z.~Liu, and M.~Zhou, ``Fast and accurate non-negative latent factor analysis of high-dimensional and sparse matrices in recommender systems,'' \emph{{IEEE} Trans. Knowl. Data Eng.}, vol.~35, no.~4, pp. 3897--3911, 2023.

\bibitem{LLMisCompression}
G.~Del{\'e}tang, A.~Ruoss, P.-A. Duquenne, E.~Catt, T.~Genewein, C.~Mattern, J.~Grau-Moya, L.~K. Wenliang, M.~Aitchison, L.~Orseau \emph{et~al.}, ``Language modeling is compression,'' \emph{arXiv:2309.10668}, 2023.

\bibitem{acl21best}
J.~Xu, H.~Zhou, C.~Gan, Z.~Zheng, and L.~Li, ``Vocabulary learning via optimal transport for neural machine translation,'' in \emph{{ACL/IJCNLP} 2021}, 2021, pp. 7361--7373.

\bibitem{lightgcn}
X.~He, K.~Deng, X.~Wang, Y.~Li, Y.~Zhang, and M.~Wang, ``Lightgcn: Simplifying and powering graph convolution network for recommendation,'' in \emph{Proceedings of the 43rd International ACM SIGIR conference on research and development in Information Retrieval}, 2020, pp. 639--648.

\bibitem{llmrecsurvey2}
J.~Lin, X.~Dai, Y.~Xi, W.~Liu, B.~Chen, X.~Li, C.~Zhu, H.~Guo, Y.~Yu, R.~Tang \emph{et~al.}, ``How can recommender systems benefit from large language models: A survey,'' \emph{arXiv:2306.05817}, 2023.

\bibitem{ai2023information}
Q.~Ai, T.~Bai, Z.~Cao, Y.~Chang, J.~Chen, Z.~Chen, Z.~Cheng, S.~Dong, Z.~Dou, F.~Feng \emph{et~al.}, ``Information retrieval meets large language models: A strategic report from chinese ir community,'' \emph{AI Open}, vol.~4, pp. 80--90, 2023.

\bibitem{ChatgptGoodRec}
J.~Liu, C.~Liu, R.~Lv, K.~Zhou, and Y.~Zhang, ``Is chatgpt a good recommender? a preliminary study,'' \emph{arXiv:2304.10149}, 2023.

\bibitem{llmconversationrec}
Y.~Feng, S.~Liu, Z.~Xue, Q.~Cai, L.~Hu, P.~Jiang, K.~Gai, and F.~Sun, ``A large language model enhanced conversational recommender system,'' \emph{CoRR}, vol. abs/2308.06212, 2023.

\bibitem{rectool}
X.~Huang, J.~Lian, Y.~Lei, J.~Yao, D.~Lian, and X.~Xie, ``Recommender ai agent: Integrating large language models for interactive recommendations,'' \emph{arXiv:2308.16505}, 2023.

\bibitem{icldiff}
J.~Wei, J.~Wei, Y.~Tay, D.~Tran, A.~Webson, Y.~Lu, X.~Chen, H.~Liu, D.~Huang, D.~Zhou \emph{et~al.}, ``Larger language models do in-context learning differently,'' \emph{arXiv:2303.03846}, 2023.

\bibitem{jundong}
Y.~Zhu, L.~Wu, Q.~Guo, L.~Hong, and J.~Li, ``Collaborative large language model for recommender systems,'' in \emph{Proceedings of the ACM on Web Conference 2024}, 2024, pp. 3162--3172.

\bibitem{lc-rec}
B.~Zheng, Y.~Hou, H.~Lu, Y.~Chen, W.~X. Zhao, M.~Chen, and J.-R. Wen, ``Adapting large language models by integrating collaborative semantics for recommendation,'' in \emph{2024 IEEE 40th International Conference on Data Engineering (ICDE)}.\hskip 1em plus 0.5em minus 0.4em\relax IEEE, 2024, pp. 1435--1448.

\bibitem{ctr_bert}
A.~Muhamed, I.~Keivanloo, S.~Perera, J.~Mracek, Y.~Xu, Q.~Cui, S.~Rajagopalan, B.~Zeng, and T.~Chilimbi, ``Ctr-bert: Cost-effective knowledge distillation for billion-parameter teacher models,'' in \emph{NeurIPS Efficient Natural Language and Speech Processing Workshop}, 2021.

\bibitem{vision1}
S.~Wu, H.~Fei, L.~Qu, W.~Ji, and T.-S. Chua, ``Next-gpt: Any-to-any multimodal llm,'' \emph{arXiv:2309.05519}, 2023.

\bibitem{minigpt4}
D.~Zhu, J.~Chen, X.~Shen, X.~Li, and M.~Elhoseiny, ``Mini{GPT}-4: Enhancing vision-language understanding with advanced large language models,'' in \emph{The Twelfth International Conference on Learning Representations}, 2024.

\bibitem{speech1}
D.~Zhang, S.~Li, X.~Zhang, J.~Zhan, P.~Wang, Y.~Zhou, and X.~Qiu, ``Speech{GPT}: Empowering large language models with intrinsic cross-modal conversational abilities,'' in \emph{EMNLP}, 2023.

\bibitem{palme}
D.~Driess, F.~Xia, M.~S.~M. Sajjadi, C.~Lynch, A.~Chowdhery, B.~Ichter, A.~Wahid, J.~Tompson, Q.~Vuong, T.~Yu, W.~Huang, Y.~Chebotar, P.~Sermanet, D.~Duckworth, S.~Levine, V.~Vanhoucke, K.~Hausman, M.~Toussaint, K.~Greff, A.~Zeng, I.~Mordatch, and P.~Florence, ``Palm-e: an embodied multimodal language model,'' in \emph{Proceedings of the 40th International Conference on Machine Learning}, 2023.

\bibitem{vicuna2023}
\BIBentryALTinterwordspacing
W.-L. Chiang, Z.~Li, Z.~Lin, Y.~Sheng, Z.~Wu, H.~Zhang, L.~Zheng, S.~Zhuang, Y.~Zhuang, J.~E. Gonzalez, I.~Stoica, and E.~P. Xing, ``Vicuna: An open-source chatbot impressing gpt-4 with 90\%* chatgpt quality,'' March 2023. [Online]. Available: \url{https://lmsys.org/blog/2023-03-30-vicuna/}
\BIBentrySTDinterwordspacing

\bibitem{lora}
E.~J. Hu, Y.~Shen, P.~Wallis, Z.~Allen{-}Zhu, Y.~Li, S.~Wang, L.~Wang, and W.~Chen, ``Lora: Low-rank adaptation of large language models,'' in \emph{ICLR}, 2022.

\bibitem{gim2023prompt}
I.~Gim, G.~Chen, S.-s. Lee, N.~Sarda, A.~Khandelwal, and L.~Zhong, ``Prompt cache: Modular attention reuse for low-latency inference,'' \emph{arXiv:2311.04934}, 2023.

\bibitem{movielendata}
F.~M. Harper and J.~A. Konstan, ``The movielens datasets: History and context,'' \emph{Acm transactions on interactive intelligent systems (tiis)}, vol.~5, no.~4, pp. 1--19, 2015.

\bibitem{wenjieDiffusion}
W.~Wang, Y.~Xu, F.~Feng, X.~Lin, X.~He, and T.-S. Chua, ``Diffusion recommender model,'' in \emph{SIGIR}, 2023, p. 832–841.

\bibitem{data_Leakage_1}
Y.~Ji, A.~Sun, J.~Zhang, and C.~Li, ``A critical study on data leakage in recommender system offline evaluation,'' \emph{{ACM} Trans. Inf. Syst.}, vol.~41, no.~3, pp. 75:1--75:27, 2023.

\bibitem{zhang2020retrain}
Y.~Zhang, F.~Feng, C.~Wang, X.~He, M.~Wang, Y.~Li, and Y.~Zhang, ``How to retrain recommender system? a sequential meta-learning method,'' in \emph{Proceedings of the 43rd International ACM SIGIR Conference on Research and Development in Information Retrieval}, 2020, pp. 1479--1488.

\bibitem{auc-uauc}
Y.~Liu, Q.~Liu, Y.~Tian, C.~Wang, Y.~Niu, Y.~Song, and C.~Li, ``Concept-aware denoising graph neural network for micro-video recommendation,'' in \emph{Proceedings of the 30th ACM International Conference on Information \& Knowledge Management}, 2021, pp. 1099--1108.

\bibitem{ndcg}
K.~J{\"a}rvelin and J.~Kek{\"a}l{\"a}inen, ``Cumulated gain-based evaluation of ir techniques,'' \emph{ACM Transactions on Information Systems (TOIS)}, vol.~20, no.~4, pp. 422--446, 2002.

\bibitem{SASRec}
W.-C. Kang and J.~McAuley, ``Self-attentive sequential recommendation,'' in \emph{ICDM}, 2018, pp. 197--206.

\bibitem{DIN}
G.~Zhou, X.~Zhu, C.~Song, Y.~Fan, H.~Zhu, X.~Ma, Y.~Yan, J.~Jin, H.~Li, and K.~Gai, ``Deep interest network for click-through rate prediction,'' in \emph{Proceedings of the 24th ACM SIGKDD international conference on knowledge discovery \& data mining}, 2018, pp. 1059--1068.

\bibitem{prompt4nr}
Z.~Zhang and B.~Wang, ``Prompt learning for news recommendation,'' in \emph{Proceedings of the 46th International {ACM} {SIGIR} Conference on Research and Development in Information Retrieval}.\hskip 1em plus 0.5em minus 0.4em\relax {ACM}, 2023, pp. 227--237.

\bibitem{bert}
J.~Devlin, M.~Chang, K.~Lee, and K.~Toutanova, ``{BERT:} pre-training of deep bidirectional transformers for language understanding,'' in \emph{NAACL-HLT}, 2019, pp. 4171--4186.

\bibitem{p5}
S.~Geng, S.~Liu, Z.~Fu, Y.~Ge, and Y.~Zhang, ``Recommendation as language processing (rlp): A unified pretrain, personalized prompt \& predict paradigm (p5),'' in \emph{Proceedings of the 16th ACM Conference on Recommender Systems}, 2022, pp. 299--315.

\bibitem{adam}
D.~P. Kingma and J.~Ba, ``Adam: {A} method for stochastic optimization,'' in \emph{ICLR}, 2015.

\bibitem{ensemble}
T.~G. Dietterich, ``Ensemble methods in machine learning,'' in \emph{International workshop on multiple classifier systems}.\hskip 1em plus 0.5em minus 0.4em\relax Springer, 2000, pp. 1--15.

\bibitem{9429954}
Y.~Xu, L.~Zhu, Z.~Cheng, J.~Li, Z.~Zhang, and H.~Zhang, ``Multi-modal discrete collaborative filtering for efficient cold-start recommendation,'' \emph{IEEE Transactions on Knowledge and Data Engineering}, vol.~35, no.~1, pp. 741--755, 2023.

\bibitem{yang2024qwen2}
A.~Yang, B.~Yang, B.~Hui, B.~Zheng, B.~Yu, C.~Zhou, C.~Li, C.~Li, D.~Liu, F.~Huang \emph{et~al.}, ``Qwen2 technical report,'' \emph{arXiv preprint arXiv:2407.10671}, 2024.

\end{thebibliography}

\vspace{-1cm}
\begin{IEEEbiographynophoto}
{Yang Zhang} is a Research Fellow at the National University of Singapore. He obtained his PhD from the University of Science and Technology of China (USTC). His research interest lies in the recommender system.
He has authored over ten publications featured in top conferences and journals like SIGIR and TKDE, with one winning the Best Paper Honorable Mention at SIGIR 2021. He has served as the program committee member, and reviewer for top conferences/journals, including KDD, SIGIR, TKDE, TOIS, etc.
\end{IEEEbiographynophoto}

\vspace{-1.3cm}
\begin{IEEEbiographynophoto}
{Fuli Feng} is a professor at the University of Science and Technology of China (USTC). His research interests include information retrieval and data mining. He has about 100 publications that appeared in top conferences such as SIGIR, WWW, and journals including TKDE and TOIS. He received the Best Paper Honourable Mention in SIGIR 2021 and the Best Poster Award in WWW 2018. Moreover, He has served as the Associate Editor for ACM TORS, the SPC/PC-member for top-tier conferences including SIGIR, WWW, SIGKDD, NeurIPS, ICLR, ICML, ACL, and the invited reviewer for prestigious journals such as TOIS, TKDE, TPAMI, TNNLS. 
\end{IEEEbiographynophoto}

\vspace{-1cm}
\begin{IEEEbiographynophoto}
{Jizhi Zhang} is a Ph.D. student at the University of Science and Technology of China (USTC), supervised by Prof. Fuli Feng and Prof. Xiangnan He. His research interest lies in the recommender system and LLMs. He has several publications in top conferences such as SIGIR, RecSys, and ACL. He also serves as a reviewer for academic journals including TOIS and TORS.
\end{IEEEbiographynophoto}

\vspace{-1.4cm}
\begin{IEEEbiographynophoto}
{Keqin Bao} is a Ph.D. student at University of Science and Technology of China (USTC), supervised by Prof. Fuli Feng and Prof. Xiangnan He. His research interest lies in the recommender system and LLMs. He has several publications in top conferences such as RecSys, EMNLP and WWW. He has served as the PC member and reviewer for the top conferences and journals including TOIS and RecSys. 
\end{IEEEbiographynophoto}

\vspace{-1.4cm}
\begin{IEEEbiographynophoto}
{Qifan Wang} is a Research Scientist at Meta AI, leading a team building innovative Deep Learning and Natural Language Processing models for Recommendation System. He received his PhD in computer science from Purdue University in 2015. His research interests include deep learning, natural language processing, information retrieval, data mining, and computer vision. He has co-authored over 100 publications in top-tier conferences and journals, including NeurIPS, ICLR, ICML, ACL, CVPR, SIGKDD, WWW, SIGIR, TPAMI, TKDE, TOIS, etc. He also serves as area chair, program committee member, editorial board member, and reviewer for academic conferences and journals.
\end{IEEEbiographynophoto}

\vspace{-1.2cm}

\begin{IEEEbiographynophoto}
{Xiangnan He} is a Professor at the University of Science and Technology of China (USTC). 
His research interests span recommender system, data mining, and multi-media analytics.
He has over 100 publications that have appeared in top conferences such as SIGIR, WWW, and KDD, and journals including TKDE and TOIS. His work on recommender systems has received the Best Paper Award Honorable Mention in SIGIR 2023/2021/2016 and WWW 2018. He has served as the Associate Editor for journals including ACM TOIS, IEEE TKDE, etc.
Moreover, he has served as SPC/PC member
for several top conferences including SIGIR, WWW, KDD, MM, WSDM,
ICML etc., and the regular reviewer for journals including TKDE, TOIS,
etc
and (senior) PC member for conferences including SIGIR, WWW, KDD, MM, etc. He is a senior member of IEEE.
\end{IEEEbiographynophoto}
\end{document}